\documentclass[prl,superscriptaddress,twocolumn,letterpaper]{revtex4}
 
\usepackage{amsmath}
\usepackage{bm}
\usepackage{graphicx}

\newcommand{\cuo}{CuO$_2$}
\newcommand{\bp}{\ensuremath{\mathbf{p}}}
\newcommand{\eps}{\ensuremath{\varepsilon}}
\newcommand{\Eref}[1]{Eq.~(\ref{#1})}
\newcommand{\Fref}[1]{Fig.~\ref{#1}}

\begin{document}

\title{The $3d$-to-$4s$-by-$2p$ highway to superconductivity in cuprates}

\author{T.~M.~Mishonov} 
\affiliation{Laboratorium voor Vaste-Stoffysica en Magnetisme,
Katholieke Universiteit Leuven, Celestijnenlaan 200~D, B-3001 Leuven,
Belgium}
\affiliation{Faculty of Physics, Sofia University ``St. Kliment
Ohridski'', 5~J.~Bourchier~Blvd., 1164~Sofia, Bulgaria}
\author{J.~O.~Indekeu}
\affiliation{Laboratorium voor Vaste-Stoffysica en Magnetisme,
Katholieke Universiteit Leuven, Celestijnenlaan 200~D, B-3001 Leuven,
Belgium}
\author{E.~S.~Penev}
\affiliation{Faculty of Physics, Sofia University ``St. Kliment
Ohridski'', 5~J.~Bourchier~Blvd., 1164~Sofia, Bulgaria}

\date{\today}

\begin{abstract}
High-temperature superconductors are nowadays found in great variety
and hold technological promise. It is still an unsolved mystery that
the critical temperature $T_c$ of the basic cuprates is so high. The
answer might well be hidden in a conventional corner of theoretical
physics, overlooked in the recent hunt for exotic explanations of new
effects in these materials.  A forgotten intra-atomic $s$-$d$
two-electron exchange in the Cu atom is found to provide a strong
($\sim$~eV) electron pairing interaction. A Bardeen-Cooper-Schrieffer
approach can explain the main experimental observations and predict
the correct $d_{x^2- y^2}$ symmetry of the gap.
\end{abstract}


\maketitle

The discovery of high-temperature superconductivity~\cite{Bednorz:86}
in cuprates and the subsequent ``research rush'' have led to the
appearance of about 100,000 papers to
date~\cite{Ginzburg:00}. Virtually every fundamental process known in
condensed matter physics was probed as a possible mechanism of this
phenomenon. Nevertheless, none of the theoretical efforts resulted in
a coherent picture~\cite{Ginzburg:00}. For the conventional
superconductors the mechanism was known to be the interaction between
electrons and crystal-lattice vibrations, but the development of its
theory lagged behind the experimental findings.  The case of cuprate
high-$T_c$ superconductivity appears to be opposite: we do not
convincingly know which mechanism is to be incorporated in the
traditional Bardeen-Cooper-Schrieffer (BCS) theory~\cite{BCS:57}. Thus
the path to high-$T_c$ superconductivity in cuprates, perhaps
carefully hidden or well-forgotten, has turned into one of the
long-standing mysteries in physical science.

\begin{figure}[t]
\includegraphics[width=\columnwidth]{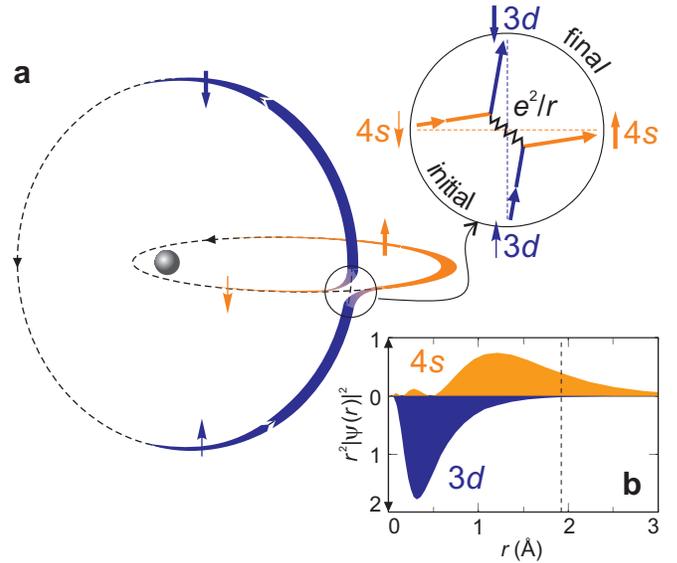}
\caption{Pairing two-electron exchange amplitude $J_{sd}$
``hidden'' in the Cu atom.  (a) Classical Bohr-Sommerfeld
representation of the $s$-$d$ two-electron exchange process. The inset
shows how the Coulomb scattering leads to an effective electron spin
exchange.  (b) Electron charge distribution for Cu~$4s$ and Cu~$3d$
orbitals: the dashed line marks the Cu-O distance in the \cuo\ plane.
\label{fig:1}
}
\end{figure}

In contrast with all previous proposals, we advance the exchange of
two electrons between the $4s$ and $3d_{x^2-y^2}$ shells of the Cu
atom as the origin of high-$T_c$ superconductivity in the layered
cuprates and show that the basic spectroscopic, thermodynamic and
kinetic experiments can be explained by it. This process has been
familiar since the dawn of quantum mechanics~\cite{Heisenberg:28} and
is known to be responsible for the magnetism of transition metals. Its
energetic characteristic is referred to as the $s$-$d$ exchange
amplitude (or integral) $J_{sd}.$ The gist of this two-electron
correlation is sketched in \Fref{fig:1}. The question then naturally
arises as to how this exchange process can trigger pairing between
electrons with opposite momenta and spins, the so-called Cooper pairs.

The underlying idea is analogous to Heitler and London's
reasoning~\cite{Heitler:27} which led to their celebrated model for
the valence bonding of the H$_2$ molecule: The strong electron
correlation brought about by the ubiquitous Coulomb repulsion lowers
the energy when the two electrons have opposite spins (the singlet
state). In quantum mechanics, according to the Hellmann-Feynman
theorem, this energy decrease ``drives'' the attractive interatomic
force. Reverting to the Cu atom, \Fref{fig:1}, one can think of the
$s$-$d$ exchange as an ``intra-atomic valence bond''. We shall further
argue that it is the specific crystal structure of the cuprates that
renders this intensive process macroscopically observable,
``disguised'' as the phenomenon of superconductivity for temperatures
lower than some critical value $T < T_c.$

\begin{figure}
\includegraphics[width=\columnwidth]{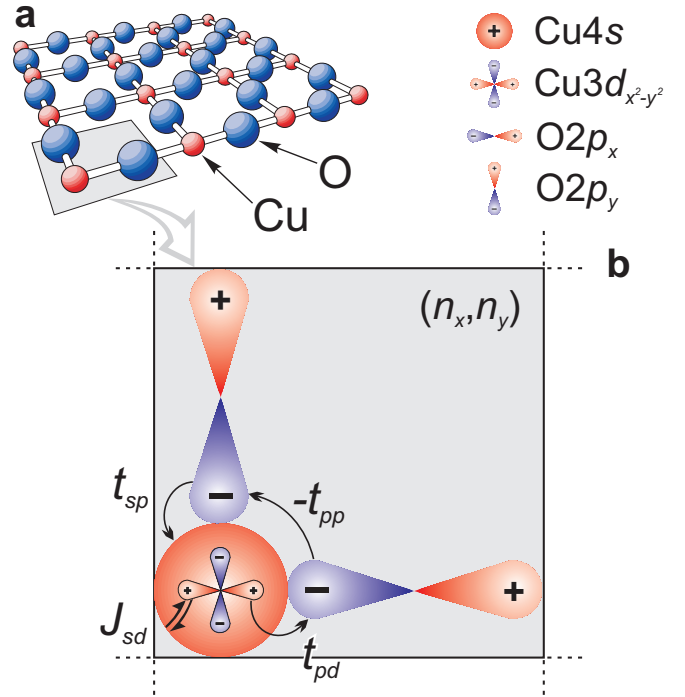}
\caption{The $3d$-to-$4s$-by-$2p$ highway in the \cuo\ plane. (a) The
shaded square is the \cuo\ unit cell indexed by
$\mathbf{n}=(n_x,n_y),\, n_{x,y}=0,\pm1,\pm2,\dots$. (b) A single
electron hops from the $3d$ atomic orbital to $2p_x$ with amplitude
$t_{pd}$, contained in $\hat{H}_0$. From $2p_x$ to $2p_y$ the
hopping amplitude is $t_{pp}$, and from there to $4s$ with amplitude
$t_{sp}$. Correlated hopping of two electrons in opposite directions
between $3d$ and $4s$ creates the pairing. The exchange integral
$J_{sd}$ from \protect\Fref{fig:1}~(a) is depicted as a double
arrow. \label{fig:2} }
\end{figure}

\begin{figure}[t]
\includegraphics[width=0.69\columnwidth]{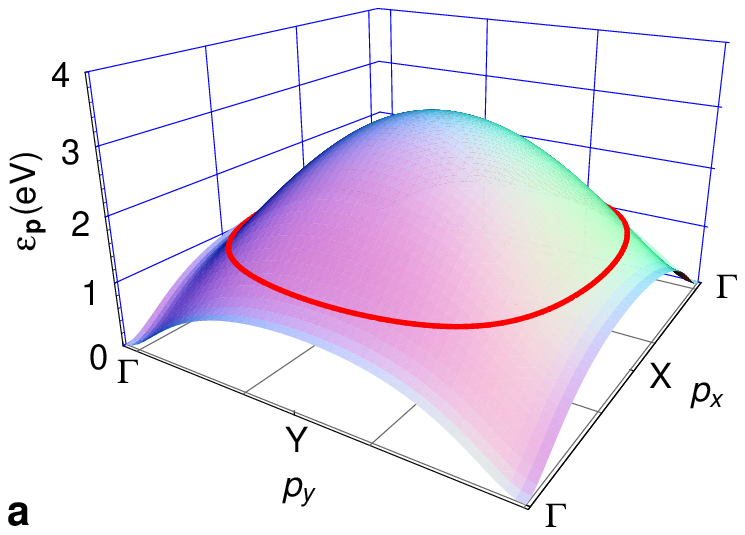}\\
\includegraphics[width=0.69\columnwidth]{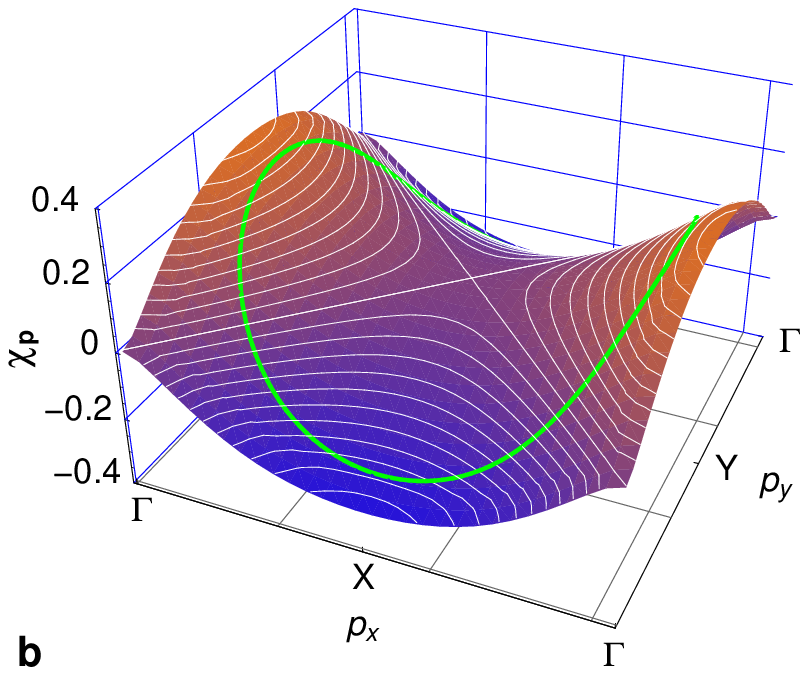}\\
\includegraphics[width=0.69\columnwidth]{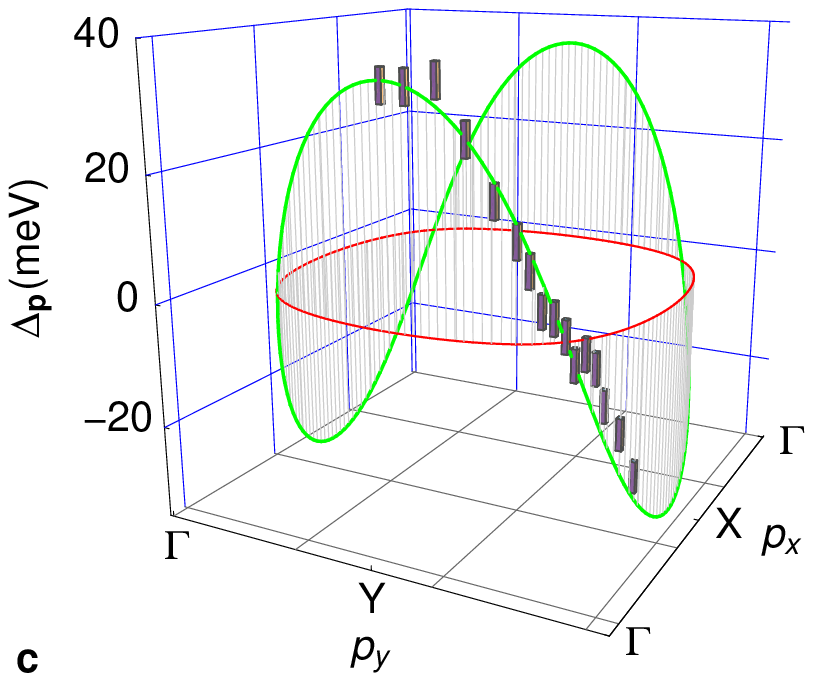}
\caption{Electronic properties of the superconducting \cuo\ plane.
(a) Conduction band energy $\eps_{\bp}$ as a function of the
quasi-momentum $\bp.$ The red contour corresponding to the Fermi
energy, $\eps_{\bp} = E_F,$ is in excellent agreement with the
ARPES data~\protect\cite{Schwaller:95}. (b) Momentum dependence of the
gap-anisotropy function $\chi_{\bp}$ within the $s$-$d$ model.  The
functional values along the Fermi contour are indicated by a green
line.  (c) Superconducting gap at zero temperature $\Delta_{\bp}$
(green line) according to our analytical result~\protect\Eref{eq:gap},
plotted along the Fermi contour (red line). The ARPES
data~\protect\cite{Mesot:99} for BSCCO are given as prisms with sizes
corresponding to the experimental error bars. The gap function along
the Fermi contour has the same qualitative behavior and symmetry as
the Cu~$3d_{x^2-y^2}$ electron wave function along the circular orbit
sketched in \protect\Fref{fig:1}~(a).
\label{fig:3} }
\end{figure}

Superconducting cuprates have as their main structural detail the
copper-oxygen plane shown in \Fref{fig:2}~(a). It has previously been
demonstrated~\cite{Mishonov:00} that its normal-phase electronic
properties can be understood on the basis of Bloch's tight-binding
formalism (chemists would call it an extended H\"uckel method) applied
to the set of atomic orbitals indicated in \Fref{fig:2}~(b).  This
plane behaves much like a two-dimensional metal. Electronic Bloch
states of quasi-momentum $\bp=(p_x,p_y) \in (0, 2\pi)$ are described
by a four-component tight-binding wave function $\psi_{\bp} =
(D_{\bp},S_{\bp},X_{\bp},Y_{\bp}).$ At $T=0$ these states are occupied
up to the Fermi energy $E_F.$ The corresponding energy level is
comprised in a single conduction band $\eps_{\bp}$ of dominant
Cu~$3d_{x^2-y^2}$ character, \Fref{fig:3}~(a). Thus, the amplitude for
a conduction electron to be Cu~$4s$ is $S_{\bp}.$ $D_{\bp}$
corresponds to Cu~$3d_{x^2- y^2},$ $X_{\bp}$ to O~$2p_x$ and $Y_{\bp}$
to O~$2p_y$, \Fref{fig:2}~(b).

The remarkable success of the \cuo\ plane in mediating high-$T_c$
superconductivity is due to the following reasons. (i) The
quasi-two-dimensional $d$-band $\eps_{\bp}$ created by $p$-$d$
hybridization is relatively narrow and the density of states rather
high. The wide $s$-band resulting mainly from $s$-$p$ hybridization is
empty, which is why up to now the $s$-$d$ model has not been
applied. (ii) In order for the $s$-$d$ exchange process to become a
pairing mechanism the $s$- and $d$-levels must be close. A virtual
population of the $s$-level by band hybridization is needed in order
to make the $J_{sd}$ amplitude operative. Indeed, the conduction
$d$-band is the result of an $s$-$p$-$d$ hybridization in the \cuo\
plane. Another favorable factor for hybridization is the proximity of
the O~$2p$ and Cu~$3d_{x^2-y^2}$ levels. As the $3d$ and $4s$ states
are orthogonal the hybridization requires a go-between, the O~$2p$
orbital. Hence this theory can be nicknamed ``$3d$-to-$4s$-by-$2p$
highway to high-$T_c$ superconductivity in cuprates''.

The foregoing discussion can be put in mathematical form. The central
entity is the many-particle Hamiltonian $\hat{H} = \hat{H}_0 + {\hat{
H}}_{\mathrm{int}}.$ Here $\hat{H}_0$ describes~\cite{Mishonov:00}
self-consistent motion of independent Fermi particles in
Bloch-H\"uckel approximation, and ${\hat{H}}_{\mathrm{int}}$ specifies
the interaction. Following the seminal works by Schubin and
Wonsowsky~\cite{Schubin:34}, Vonsovskii~\cite{Vonsovskii:46}, and
Zener~\cite{Zener:51}, cf. the Feynman lectures on
physics~\cite{Feynman:66}, we set up the $s$-$d$ pairing Hamiltonian
for the \cuo\ plane
\begin{equation}
 \hat{H}_{\mathrm{int}} = -J_{sd} \sum_{\mathbf{n},\alpha,\beta}
 \hat{S}^{\dag}_{{\bf n}\alpha}\hat{D}^{\dag}_{{\bf n}\beta}
 \hat{S}_{{\bf n}\beta}\hat{D}_{{\bf n}\alpha},\quad\!
 \alpha,\beta =\, \uparrow,\downarrow.
\label{eq:Hint}
\end{equation}

This Hamiltonian is written in the language of second quantization but
one can easily trace the connection with the classical Bohr-Sommerfeld
picture, \Fref{fig:1}~(a). $\hat{D}_{{\bf n}\alpha}$ is the Fermi
operator for annihilating a $3d_{x^2-y^2}$ electron with
spin projection $\alpha$ in the {\bf n}th unit cell. The electron is
then recreated without spin flip in the $4s$ state by the Fermi
creation operator $\hat{S}^{\dag}_{{\bf n}\alpha}$.  Simultaneously,
another electron with spin $\beta$ makes a reverse hop, from $4s$ to
$3d_{x^2-y^2}$ in the same Cu atom. Thus, the
$\hat{S}^{\dag}\hat{D}^{\dag}\hat{S}\hat{D}$ product in
Eq.~\ref{eq:Hint} represents the two-electron exchange drawn as a
double arrow in \Fref{fig:2}~(b) and shown in \Fref{fig:1}~(a).

A model which incorporates a Hamiltonian of the form $\hat{H} =
\hat{H}_0 + {\hat{H}}_{\mathrm{int}}$ implies the validity of the
Fermi-liquid picture for the cuprates. That is, we assert that an
electron is merely a Fermi quasiparticle and not a composite object
displaying spin-charge separation. Two key
experiments~\cite{Feng:01,Proust:02} have shown that this is
correct. The long-sought bilayer splitting in the electronic structure
of the cuprates was observed~\cite{Feng:01} for overdoped
Bi$_2$Sr$_2$CaCu$_2$O$_{8+\delta}$ (BSCCO). Furthermore, the overdoped
Tl$_2$Ba$_2$CuO$_{6+\delta}$ was found~\cite{Proust:02} to obey the
150-years-old Wiedemann-Franz law to a remarkable accuracy of the
Sommerfeld's value for the Lorentz ratio.  Let us stress also that the
linear temperature dependence of the normal-state resistivity is an
intrinsic property~\cite{Mishonov2:00} of the ``layered'' electron gas
and cannot be used as an argument in favor of non-Fermi-liquid
behavior.

We now apply the traditional BCS approach~\cite{BCS:57} to the
Hamiltonian $\hat{H}$ and derive an equation for the basic
characteristic of superconductors---the energy gap $\Delta_{\bp}(T)$
(the energy needed to break up a Cooper pair is $2\Delta$).  The
local, intra-atomic character of the pairing interaction tremendously
simplifies the BCS gap equation and the solution has the plain form
\begin{equation}
 \Delta_{\bp}(T) = \Xi(T)\,\chi_{\bp},\qquad  \chi_{\bp} \equiv S_{\bp}D_{\bp}.
\label{eq:gap}
\end{equation}
The order parameter $\Xi(T),$ which satisfies the standard BCS gap
equation, is maximal at $T=0$ and vanishes for $T\rightarrow T_c.$ The
$S_{\bp}D_{\bp}$ product is the amplitude of the $s$-$d$ hybridization
of the conduction band. In this theory it determines the momentum
dependence of the superconducting gap and is shown in
\Fref{fig:3}~(b). Note that $\chi_{\bp}$ exhibits the desired $d$-wave
symmetry which is now well established
experimentally~\cite{Tsuei:00}. It is a direct consequence of the
electron band hybridization. The mediator O~$2p$ orbitals transmit the
symmetry of the Cu~$3d_{x^2 - y^2}$ orbital to the momentum dependence
of the Bloch wave amplitude $S_{\bp}$ and the superconducting gap
$\Delta_{\bp}(T)$. This is why the pairing state has $d$-wave symmetry
directly coming from the Cu~$3d_{x^2-y^2}$ orbital.

The angle-resolved photoemission spectroscopy (ARPES) provides a
straightforward method for measuring $\Delta_{\bp}$. We have fitted
our theoretical result Eq.~\ref{eq:gap} to the recent precise ARPES
data~\cite{Mesot:99} for BSCCO and the outcome is given in
\Fref{fig:3}~(c). It is the first demonstration that the theory
provides a satisfactory description of overdoped cuprates.

We mentioned some normal state
properties~\cite{Feng:01,Proust:02,Mishonov2:00} of overdoped cuprates
which are governed by conventional physics.  The point to be stressed
now is that not only the normal phase but even the superconducting
phase at optimal doping, for which the $T_c$ is the highest, can be
explained by conventional theory, since there is overwhelming evidence
for the applicability of the Landau quasiparticle picture.

The quantitative analysis of the thermal conductivity~\cite{Chiao:00}
of optimally doped Bi$_2$Sr$_2$CaCu$_2$O$_8$ gave values for the Fermi
velocity $v_F$ and derivative of the gap function with
respect to the momentum component tangential to the Fermi contour,
$v_{\Delta}=\partial \Delta/\partial p_l,$ within $10\%$ of the values
found in independent ARPES experiments, as emphasized in a recent
review~\cite{Orenstein:00}. Furthermore, measurements of heat
transport, specific heat, and superfluid density through the
penetration depth also strongly support the existence of low-energy
Fermi excitations of the conventional BCS ground state for a $d$-wave
superconductor. Moreover, the quality of the quasiparticles is
astonishing since their mean-free path in some high-quality crystals
can reach 2500~\AA\ at 15~K, as reviewed by Lee~\cite{Lee:97}.

Our calculations consist of solving the gap equation in the framework
of the BCS theory using linear combination of atomic orbitals for the
electron band structure. In contrast with theoretical explanations
advanced so far our results indicate that the extraordinarily large
pairing amplitude together with the correct gap symmetry cannot be
explained by any interatomic or even interlayer exchange
interaction. Instead it must be due to the powerful intra-atomic
two-electron exchange process, which incidentally is responsible for
the magnetism of the transition metals as well. 
The superconductivity of the optimally doped cuprates, most important
for practical applications, can now be understood. The pairing
amplitude we uncover is so strong that the high $T_c$ is robust to
many other new effects discovered in optimally doped cuprates,
including the existence of a weak pseudo-gap~\cite{Arko:99}, the
appearance of stripes~\cite{Howald:02,Cho:02}, and the confluence of
superconductivity and magnetism~\cite{Ramires:99}.
Although these new effects can be considered to make the physics of
cuprates revolutionary, and require novel theoretical ideas for their
explanation, our point is that all these effects are irrelevant to the
main issue of explaining the high $T_c$ of high-$T_c$
superconductors~\cite{Anderson:00}.

The proposed model contains energies and hopping parameters which have
to be determined by fits to experimental data or by first-principles
calculations. Predicting thermodynamic properties close to $T_c$
constitutes a stringest test of the model.  For example, let us recall
that the BCS result for the jump of the specific heat in isotropic-gap
superconductors, $\Delta C$, at $T=T_c$, relative to the specific heat
of the normal phase $C_n$, is given by $\Delta
C/C_n=12/(7\zeta(3))\approx 1.43$ ($\zeta$ is the Riemann zeta
function). For the overdoped samples we find that this ratio is
renormalized by the gap anisotropy according to the relation
\begin{equation}
\frac{\Delta C}{C_n} =1.43 \,\frac{\langle|\Delta_{\bp}|^2\rangle^2
 }{
   \langle 1 \rangle\langle|\Delta_{\bp}|^4\rangle
 }.
\end{equation}
Here $\langle\dots\rangle$ denotes averaging over the Fermi surface
and $\langle1\rangle$ is the density of states. This simple expression
provides another direct means for experimental verification of the
model's validity for overdoped samples.

As a fundamental microscopic process the two-electron $s$-$d$ exchange
has been known in the physics of magnetism long
before~\cite{Schubin:34,Vonsovskii:46,Zener:51} the advent of the BCS
theory~\cite{BCS:57}. So it is not surprising that an exchange-driven
superconductivity was attempted shortly after~\cite{Akhiezer:59} the
BCS theory. Reventuring this idea we have shown how the two paradigms
($s$-$d$ \& BCS) can be reconciled to obtain a coherent traditional
theory of high-$T_c$ superconductivity in the
cuprates. Magnetism of transition metals and high-$T_c$
superconductivity of cuprates seem to be two faces of the same
ubiquitous two-electron exchange amplitude.

\begin{acknowledgments}
This work was supported by the Flemish Programmes VIS, IUAP and GOA.
\end{acknowledgments}

\bibliographystyle{apsrev}

\end{document}